\documentclass{PoS}

\newcommand{\nn}{\nonumber\\}
\newcommand{\ba}{\begin{eqnarray}}
\newcommand{\ea}{\end{eqnarray}}
\newcommand{\la}[1]{\label{#1}}

\title{30 Years of Jet Quenching}

\ShortTitle{30 years of jet quenching}

\author{Xin-Nian Wang
\\
Key Laboratory of Quark \& Lepton Physics (MOE) and Institute of Particle Physics, Central China Normal University, Wuhan 430079, China\\
Nuclear Science Division, Lawrence Berkeley National Laboratory, Berkeley, CA 94720, USA \\
        E-mail: \email{xnwang@lbl.gov}}

\abstract{In the last 30 years, the physics of jet quenching has gone from an early stage of a pure theoretical idea to initial theoretical calculations, experimental verification and now a powerful diagnostic tool for studying properties of the quark-gluon plasma (QGP) in high-energy heavy-ion collisions. I will describe my collaboration with Miklos Gyulassy in this exciting area of high-energy nuclear physics in the past 30 years on this special occasion of his 70th birthday and discuss what is ahead of us in jet tomographic study of QGP in heavy-ion collisions.}

\FullConference{13th International Workshop on High-pT Physics in the RHIC/LHC Era\\
		19-22 March 2019\\
		Symposium on the Past, Present and Future of Relativistic Heavy-ion Collisions,\\
		 in honor of Miklos Gyulassy's 70th Birthday\\
		18 March 2019\\
		University of Tennessee, Knoxville, Tennessee, USA}

\begin{document}

\section{Early years of jet quenching studies}

30 years ago back in 1989, several events occurred that have eventual consequences in the rest of my life and shaped both my scientific career and my view of the world. Around June 4, 1989, the student movement at the Tiananmen Square in Beijing ended tragically in front of my TV screen while I was studying at University of Oregon. Even though half globe  away from the epic center of the movement, I became deeply attached to those students since I knew I would be among them there if I had not come to US to study for my Ph.D. 4 four years ago.  Amid this turmoil and tragedy in the remote homeland, I still managed to publish three papers in Physical Review D on dynamics of multiple particle production in hadron-nucleus collisions  \cite{Wang:1988bw,Hwa:1988bh} and the effect of mini-jet production in multiple particle production in $p+p$ collisions \cite{Wang:1988be}. I completed my Ph.D. thesis based on these studies and was awarded a Ph.D. on September 1, 1989 from the Physics Department at the University of Oregon under supervision of Rudy Hwa. By that time, I have accepted a postdoctoral job offer from Miklos at the Lawrence Berkeley National Laboratory (LBNL) and joined the theory group on October 1, 1989. The Loma Prieta earthquake shook the parking lot underneath my feet violently in the late afternoon of October 17 that year when I was just about to go to a reception at the International House on the campus of University of California Berkeley for international students and researchers. From the Berkeley Hill, I could feel the earth rumbling under my feet and see heavy smoke from a car repair garage in downtown Berkeley and the collapsed Cypress section of interstate 880 freeway. This was definitely an earth-shaking welcome for me to the north California and it also started my life-long collaboration with Miklos.

At that time, Miklos had just became interested in the parton energy loss in QCD medium, motivated by an early unpublished paper by Bjorken on elastic parton energy loss \cite{Bjorken:1982tu}. After realizing the elastic energy loss to be very small after correcting an error in Bjorken's preprint, he and Michael Pluemer turned their attention to inelastic energy loss such as that caused by non-perturbative mechanism behind string tension $\kappa\approx 1$ GeV/fm \cite{Gyulassy:1990dk}. They also studied the sensitivity of jet suppression to a constant parton energy loss in the QGP phase of the dense matter in heavy-ion collisions \cite{Gyulassy:1990ye}.  After a brief exploration of the intermittency problem as a result of jet production in multi-particle production \cite{Wang:1990ph} after I joined the Berkeley group, Miklos and I started building the HIJING (Heavy-Ion Jet INteraction Generator) Monte Carlo model for high-energy heavy-ion collisions \cite{Wang:1991hta,Gyulassy:1994ew} which occupied my research time for most of my two years at Berkeley as a postdoctoral fellow.  These were incredibly exciting and crazy times. The Relativistic Heavy-ion Collider (RHIC) had been approved by the US Department of Energy (DOE). Experimentalists across the US were working to propose several detector systems with a lot of R\&D activities. Theorists were searching for new ideas to diagnose signals of QGP and study its properties. We often worked together during weekends. That was when Miklos introduced me to the wonderful Peets coffee in downtown Berkeley.

\begin{figure}[htbp]
\begin{center}
    \includegraphics[width=12.0cm]{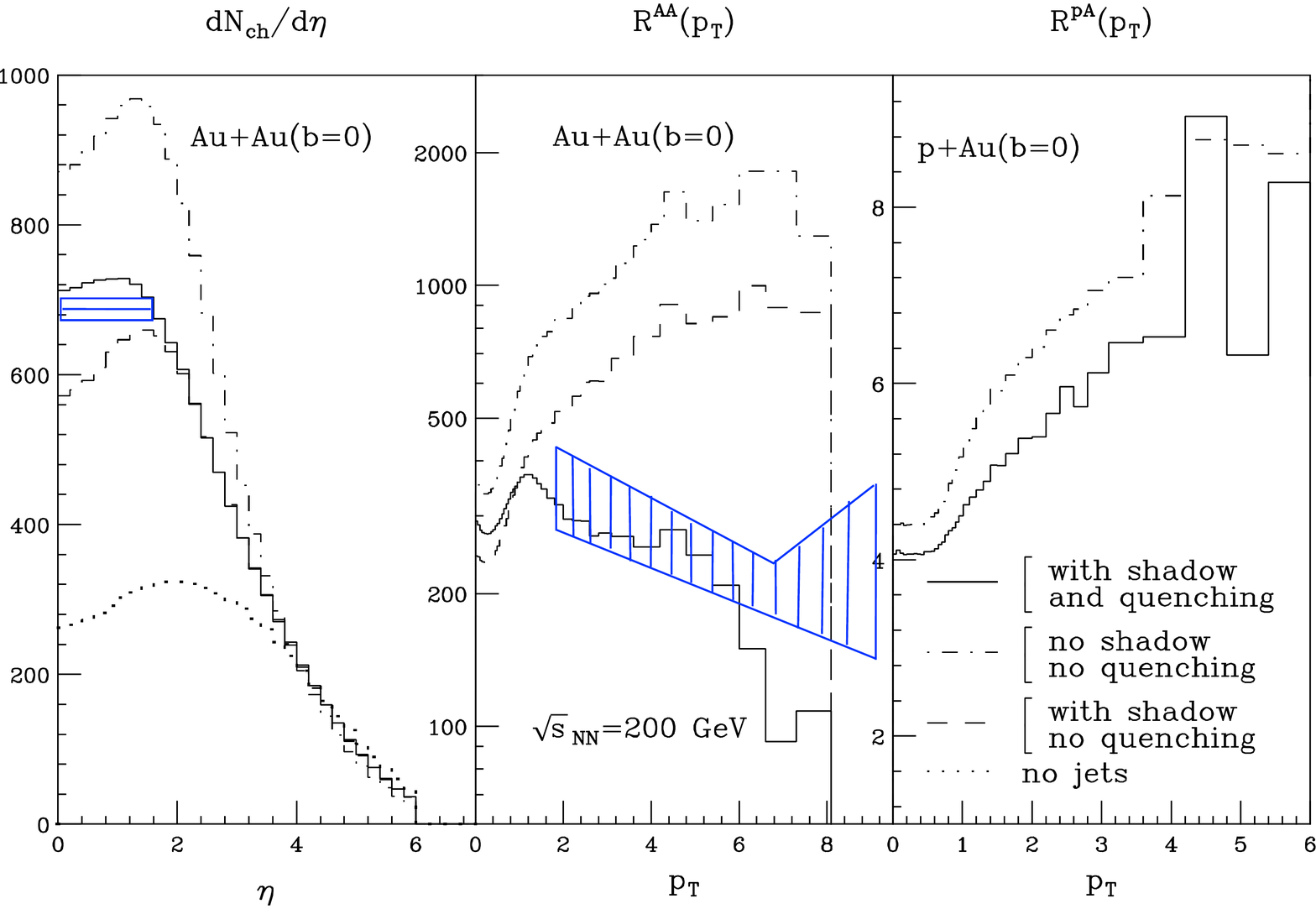}\hspace{0.2in}
    \end{center}
    \caption{Predictions for charged hadron spectra in central $Au+Au$ and $p+Au$ collisions at $\sqrt{s}=200$ GeV from HIJING \cite{Wang:1991xy} overlaid with PHENIX and STAR  results (shaded areas) at RHIC \cite{Adler:2002xw,Adler:2004zn,Adams:2003kv}.}
    \label{quenching1}
\end{figure}

HIJING model combines jet production and the associated initial and final state radiation from PYTHIA6.3 \cite{Sjostrand:1987su,Sjostrand:1986hx} with a model for multiple jet production in $p+p$, $p+A$ and $A+A$ collisions. The final state interaction between jets and dense medium was modeled by an adjustable constant parton energy loss.  An important phenomenon we discovered with HIJING is the realization that parton energy loss in medium and initial gluon shadowing can both modify the final state hadron spectra in $A+A$ collisions \cite{Wang:1991xy}.  With a default value of $dE/dx=1$ GeV/fm, the central rapidity density and the suppression of final $p_T$ spectra for charged hadrons agree with the RHIC results surprisingly well (see Fig.~\ref{quenching1}) which came more than 10 years after the publication of our HIJING prediction. Note that the suppression factor $R_{AA}$ in our first HIJING prediction in Fig.~\ref{quenching1}  was not normalized by the number of binary nucleon-nucleon scatterings. The normalized suppression factor that is commonly used nowadays was first introduced for jet quenching 10 years later  by Enke Wang and I in a study of the interplay between soft and hard processes of particle production in $p+A$ and $A+A$ collisions~\cite{Wang:2001cy}. The HIJING Monte Carlo also helped the local Relativistic Nuclear Collisions (RNC) experimental group at LBNL to explore physics capabilities of their proposed STAR (Solenoidal Tracker at RHIC) detector. Miklos and I are co-authors of the conceptual report of the STAR detector \cite{Harris:1990mh} for our efforts. It is so far our only experimental paper. This is part of the reason why HIJING was widely adopted by experimentalists for R\&D and physics simulations later on. The time between 1989 and 1991 at LBNL was two productive years for my career, not necessarily in terms of the number of publications. My daughter, Cynthia, was also born during these period, another successful story.

\section{Parton energy loss}

Around the time when we were developing HIJING in Berkeley, Miklos also brought my attention to the theoretical calculation of parton energy loss in pQCD, radiative parton energy loss in particular, since it was known at that time that elastic parton energy loss is relatively small. We continued our collaboration on parton energy loss after I moved to Duke University for my second postdoctoral job in the fall of 1991 and returned to LBNL a year later as a Divisional Fellow in the Nuclear Science Division. Miklos in the meantime also moved to the Physics Department of Columbia University in 1992. A lot of notes were faxed between us during these couple of years. 

In our first paper published on parton energy loss \cite{Gyulassy:1993hr}, we modeled multiple parton scattering in dense QCD medium with a static potential model, or the GW model, and took into account explicitly the non-Abelian Landau-Pomeranchuk-Migdal (LPM) interference for induced gluon bremsstrahlung. The radiative gluon spectrum induced by $m$ number of scatterings was expressed, 
\begin{equation}
\frac{dN_m}{d^3k}=C_m(k)\frac{dN_1}{d^3k},
\end{equation}
in terms of the gluon spectrum from a single scattering and the ``color formation factor" $C_m(k)$ which contains the color structure and LPM interference of induced gluon bremsstrahlung.  This was the first earnest attempt to calculate radiative parton energy loss with LPM effect in pQCD, even though we neglected or set aside a critical contribution from the rescattering of the radiated gluon as pointed out by the referee of our manuscript we submitted to Nuclear Physics B (we suspect the referee was one of the authors of subsequence studies on radiative parton energy loss). It motivated a flood of subsequent studies of the problem that led to the development of the foundation for modern jet tomographic study of QGP in heavy-ion collisions.

Baier, Dokshitzer, Peigne, Schiff \cite{Baier:1994bd} and later with Mueller (BDMPS) \cite{Baier:1996sk} were the first to calculate the radiative parton energy loss that took into account of rescattering of the gluonic cloud around a propagating parton and pointed out its relationship with the jet transport coefficient $\hat q$ or transverse momentum broadening squared per unit length. Zakharov subsequently formulated the problem and the LPM effect in terms of path integral \cite{Zakharov:1996fv}.  Couple of years later,  Miklos and his collaborators, Levai and Vitev (GLV) \cite{Gyulassy:1999zd} and Wiedemann \cite{Wiedemann:2000za} at Columbia University developed the framework of opacity expansion. Xiaofeng Guo and I also developed the higher-twist (HT) framework around that time to calculate the parton energy loss and the medium modification of parton fragmentation functions \cite{Guo:2000nz,Wang:2001ifa}. In this higher-twist approach, the medium modified quark fragmentation function can be obtained as a convolution,
\begin{eqnarray}
\tilde{D}_{q}^{h}(z_h,Q^2) &=&
D_{q}^{h}(z_h,Q^2)+\frac{\alpha_s(Q^2)}{2\pi}
\int_0^{Q^2}\frac{d\ell_T^2}{\ell_T^2} \int_{z_h}^{1}\frac{dz}{z} \left[ \Delta\gamma_{q\rightarrow qg}(z,x,x_L,\ell_T^2)D_{q}^h(\frac{z_h}{z})\right.
\nonumber\\
&&+ \left. \Delta\gamma_{q\rightarrow
qg}(1-z,x,x_L,\ell_T^2)D_{g}^h(\frac{z_h}{z}) \right] ,
\label{eq:mo-fragment}
\end{eqnarray}
of the vacuum fragmentation functions $D_{q,g}^h({z})$ and the medium-induced splitting function,
\begin{eqnarray}
\Delta\gamma_{q\rightarrow qg}(z,x,x_L,\ell_T^2)&=&C_A\frac{1+z^2}{(1-z)_{+}} \frac{2}{\ell_\perp^4}\int dy^- \hat q(y).
\left[1-\cos(\frac{\ell_\perp^2}{2Ez(1-z)}y^-)\right].
\end{eqnarray}
The fractional radiative energy loss is then
\begin{equation}
\frac{\Delta E}{E}=\int dz  \Delta\gamma_{q\rightarrow qg}(z,x,x_L,\ell_T^2) (1-z).
\end{equation}
This approach can be applied to quark propagation in cold nuclei and hadron suppression in semi-inclusive deeply inelastic scattering (SIDIS) off large nuclei \cite{Wang:2002ri,Chang:2014fba} as well as hadron suppression due to jet propagation in QGP in heavy-ion collisions \cite{Chen:2010te,Chen:2011vt}. In the most recent study of multiple parton scattering in SIDIS off large nuclei and Drell-Yan dilepton production in $p+A$ collisions at NLO within the higher-twist framework \cite{Kang:2013raa,Kang:2014ela,Kang:2016ron}, the factorization is verified for the transverse-momentum weighted cross section. The QCD evolution equation for twist-4 matrix elements for parton correlation or the jet transport coefficient, is also derived and its universality is demonstrated. This means that the jet transport coefficient is an intrinsic medium property independent of the hard processes that produced the hard probe or the propagating parton.

Within the framework of finite temperature QCD field theory and hard thermal loop resummation, Arnold, Moore and Yaffe (AMY) \cite{Arnold:2002ja} also studied induced gluon emission by a propagating parton in a hot QCD medium.  Together with BDMPS-Z, GLV and HT, these four different approaches formed the foundation for many theoretical and phenomenological studies of jet quenching in high-energy heavy-ion collisions \cite{Qin:2015srf}.
In the latest SCET\textsubscript{G} formalism \cite{Ovanesyan:2011xy}, the soft collinear effective theory (SCET) is supplemented with Glauber modes of soft gluon exchange for interaction between a propagating parton and static scattering centers to study multiple parton scattering and medium-induced gluon splitting. The connections between some of the above different approaches to parton energy loss have been discussed in detail in Refs.~\cite{Arnold:2008iy,CaronHuot:2010bp,Mehtar-Tani:2019tvy} and numerically compared in Ref.~\cite{Armesto:2011ht}.  Furthermore, GLV formalism has been extended beyond soft radiation approximation \cite{Blagojevic:2018nve} and with a dynamic medium through the hard thermal loop resummed gluon propagator \cite{Djordjevic:2008iz} and beyond first order in opacity expansion \cite{Sievert:2019cwq}. The HT approach has been extended to include longitudinal momentum diffusion \cite{Majumder:2009ge,Qin:2014mya} and to a generalized higher-twist framework where there is no expansion in the transverse momentum from medium exchange \cite{Zhang:2018kkn,Zhang:2018nie,Zhang:2019toi}.  Further improvements such as effects of color (de)coherence, angular order \cite{MehtarTani:2011tz,Armesto:2011ir,Caucal:2018dla}  and overlapping formation time in sequential gluon emissions \cite{Arnold:2015qya} have also been studied recently.

\section{Jet quenching phenomenology}

In high-energy heavy-ion collisions,  parton energy loss in QGP can lead to jet quenching phenomena such as the suppression of single inclusive hadron spectra, dihadron and $\gamma$-hadron correlation relative to the baseline spectra in $p+p$ collisions. Such suppressions are direct consequences of the suppression of leading hadrons from fragmentation of jets that have experienced both elastic and inelastic energy loss. This suppression of leading hadrons can be described by a medium modified jet fragmentation function. These phenomena can be used in turn as a diagnostic tool to study medium properties of QGP such as that characterized by jet transport parameters $\hat q$.  My early study of jet quenching phenomenology before first RHIC experimental data focused mainly on the medium modification of the $\gamma$-hadron correlation \cite{Wang:1996yh,Wang:1996pe}, single hadron spectra \cite{Wang:1998bha,Wang:1998hs,Wang:1998ww,Wang:2001cy} and anisotropy of high $p_T$ hadron spectra \cite{Wang:2000fq} due to path-length dependence of jet quenching in non-central heavy-ion collisions.

\begin{figure}[htbp]
\begin{center}
    \includegraphics[width=10.0cm]{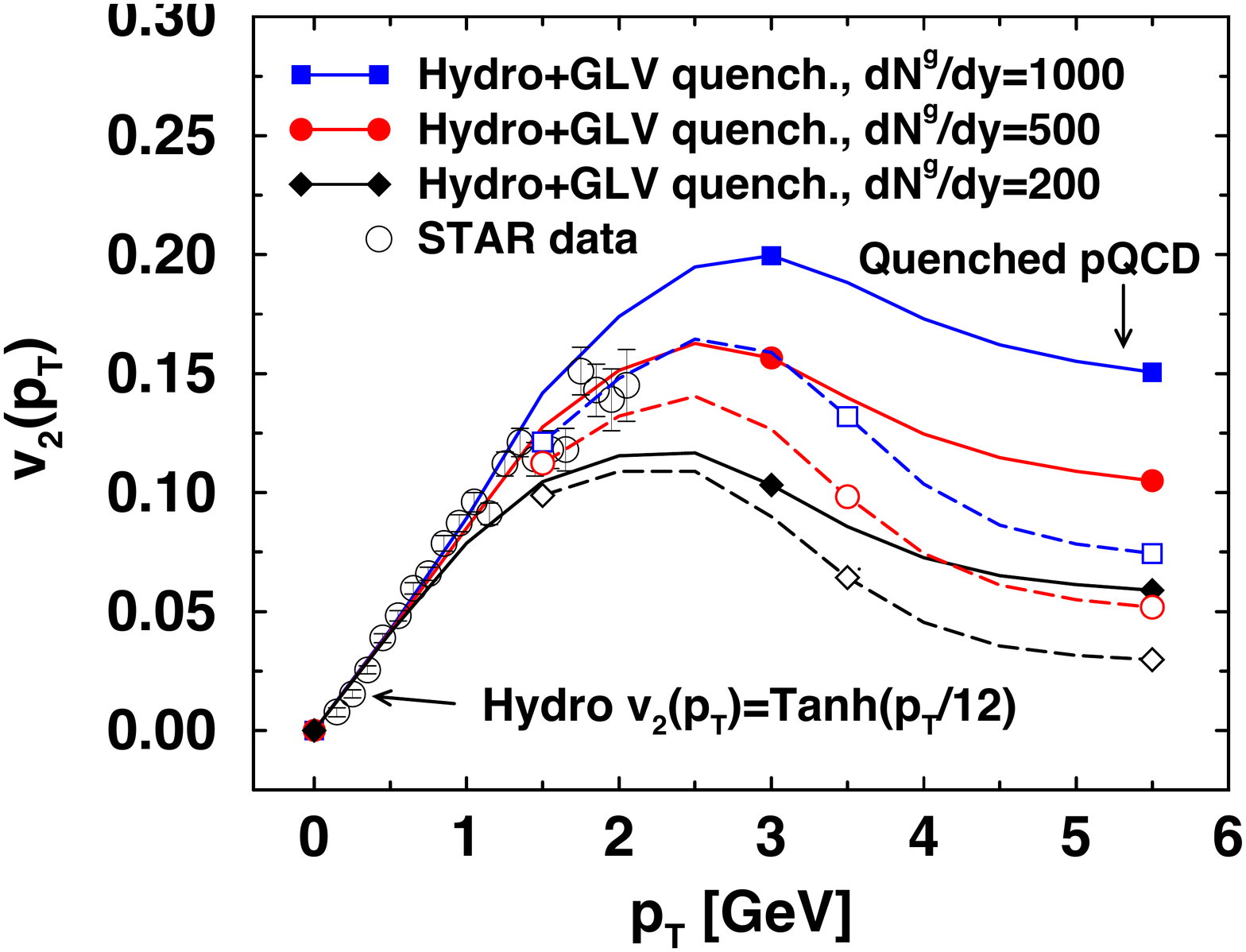}\hspace{0.2in}
    \end{center}
    \caption{Interpolation of $v_2(p_T)$ between soft hydrodynamic and hard hadron production in semi-central $Au+Au$ collisions at RHIC. See Ref.~\cite{Gyulassy:2000gk} for a detailed explanation.}
    \label{v2}
\end{figure}

Miklos, Vitev, Huovinen and I collaborated on the study of high $p_T$ hadron azimuthal anisotropy in non-central $A+A$ collisions due to the length dependence of the parton energy loss  \cite{Gyulassy:2000gk} and the influence of transverse expansion \cite{Gyulassy:2001kr} which was found to reduce the azimuthal anisotropy of the parton energy loss. We predicted that hadron elliptic flow $v_2$ will fall off from their peak values due to hydrodynamic expansion and decrease with $p_T$ at large transverse momentum due to the interplay between soft and hard hadron production mechanism and azimuthal anisotropy due to path-length dependence of the parton energy loss (see Fig.~\ref{v2}). Currently quantitative explanation of large $v_2$ at intermediate $p_T=2-10$ GeV/$c$ still remains a puzzle. Miklos and his collaborators have proposed a model of semi-quark-gluon monopole plasma to explain this within the GLV formalism for parton energy loss \cite{Xu:2014tda,Shi:2018vys}.
  
  \begin{figure}
\centering
\includegraphics[width=3.5in]{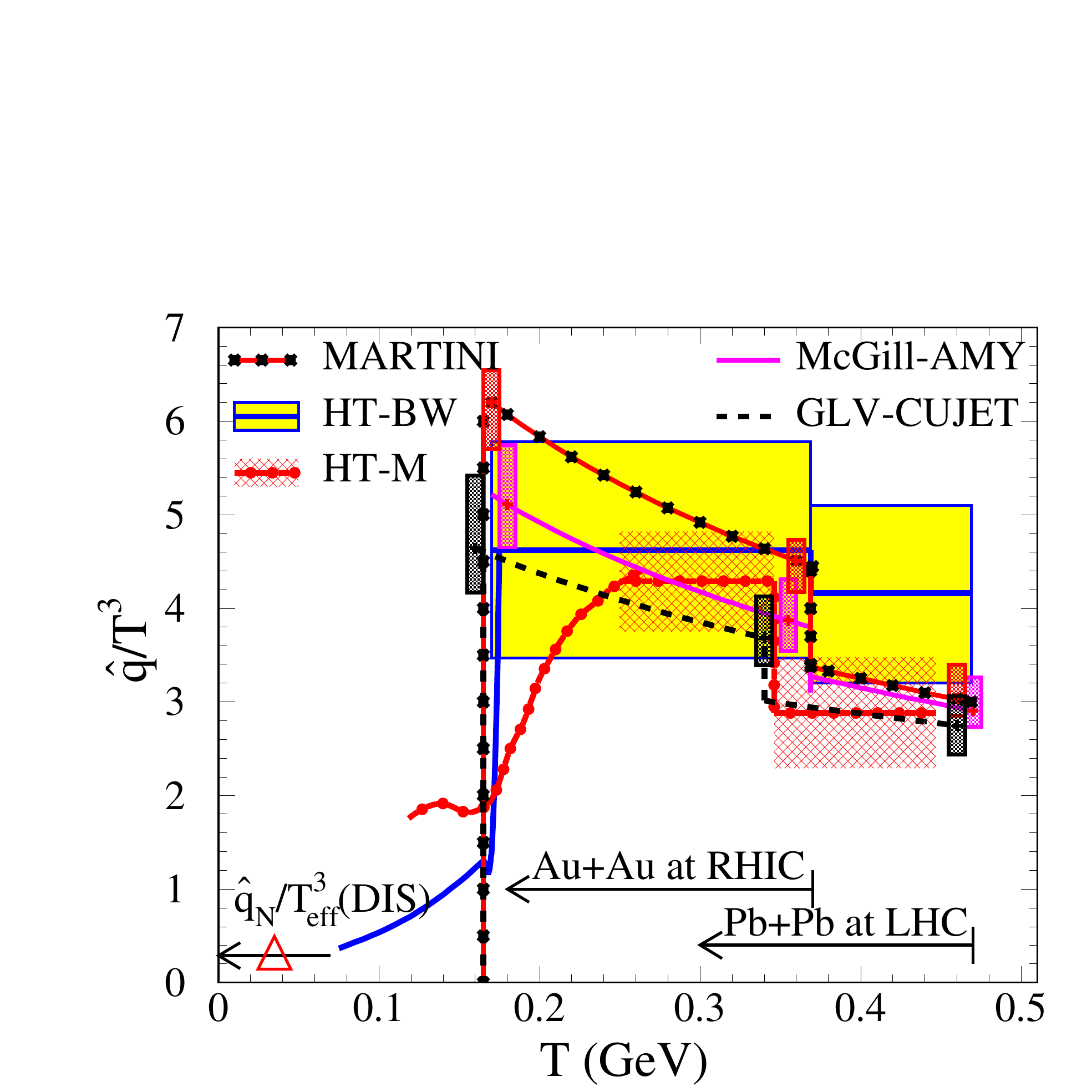}
\caption{\label{fig:raa_ht} The scaled jet transport parameter $\hat q/T^3$ in different jet quenching models for an initial quark jet with energy $E=10$ GeV as extracted from fitting to experimental data on hadron suppression factor $R_{AA}$ at both RHIC and LHC by the JET Collaboration. See Ref.~\cite{Burke:2013yra} for detailed explanation.} 
\label{qhat}
\end{figure}

My collaboration with Miklos continued within the JET (Jet and Electromagnetic Tomography) Collaboration which was funded in 2010 by DOE to extend the theoretical framework for jet-medium interaction and reduce uncertainties intrinsic to theoretical studies; develop new and powerful Monte Carlo algorithms for jet propagation and evolution inside a dynamic medium;  implement in the jet-medium interaction a realistic space-time evolution of the bulk medium as described by a combination of viscous hydrodynamics with parton and hadron cascades; and carry out systematic phenomenological studies of experimental data on single hadron spectra. Towards the end of the funding period for the JET Collaboration, we carried out a systematic and phenomenological study of experimental data on suppression of large $p_T$ single inclusive hadrons in heavy-ion collisions at both RHIC and the Large Hadron Collider (LHC) within five different approaches to parton propagation and energy loss in dense matter.  The evolution of bulk medium used in this study was given by 2+1D or 3+1D hydrodynamic models which are also constrained by experimental data on bulk hadron spectra. This latest phenomenological study by the JET Collaboration \cite{Burke:2013yra} has provided the most stringent constraints on the jet transport parameters $\hat q$ which is used to characterize the interaction between jet and medium partons as shown in Fig.~\ref{qhat}. For a quark with initial energy of 10 GeV we find that $\hat q\approx 1.2 \pm 0.3$ GeV$^2$/fm at an initial time $\tau_0=0.6$ fm/$c$ in $Au+Au$ collisions at $\sqrt{s}=200$ GeV/n and $\hat q\approx 1.9 \pm 0.7 $ GeV$^2$/fm in Pb+Pb collisions at $\sqrt{s}=2.76 $ TeV/n.

\section{Linear Boltzmann Transport model}

Medium modification of the production cross section and the structure of fully reconstructed jets in high-energy heavy-ion collisions can provide additional constraints on jet-medium interaction as first pointed out by Vitev and Zhang \cite{Vitev:2008rz}.  Since jets are collimated clusters of hadrons within a given jet cone, they are determined  not only by energy loss of leading partons but also by how the lost energy is propagated in the medium through radiated gluons and recoil medium partons. It is therefore necessary to include recoil partons and their propagation in the form of jet-induced medium response in the study of full jet suppression and medium modification.  

Starting in around 2008, my student collaborators in Central China Normal University (CCNU) and I have developed the linear Boltzmann transport (LBT) model \cite{Li:2010ts}  to study jet transport in hot QGP medium with a special emphasis on the effect of thermal recoil partons and their further transport through the medium in the form of jet-induced medium response. The transport of both jet shower and recoil partons in the LBT model is described by the linear Boltzmann equations,
\begin{eqnarray}
p_a\cdot\partial f_a&=&\int \sum_{b c d } \prod_{i=b,c,d}\frac{d^3p_i}{2E_i(2\pi)^3} (f_cf_d-f_af_b)|{\cal M}_{ab\rightarrow cd}|^2
\nn && \hspace{-0.5in}\times \frac{\gamma_b}{2}
S_2(\hat s,\hat t,\hat u)(2\pi)^4\delta^4(p_a\!+\!p_b\!-\!p_c\!-\!p_d)+ {\rm inelastic},
\label{bteq}
\end{eqnarray}
where the summation is over all possible parton flavors and scattering channels, $f_i=(2\pi)^3\delta^3(\vec{p}-\vec{p_i})\delta^3(\vec{x}-\vec{x_i}-\vec{v_i}t)$ $(i=a,c)$ are the phase-space density for jet shower partons before and after scattering and medium recoil partons,  $f_i=1/(e^{p_i\cdot u/T}\pm1)$ $(i=b,d)$ are phase-space distributions for thermal partons in the QGP medium with local temperature $T$ and fluid velocity $u=(1, \vec{v})/\sqrt{1-\vec{v}^2}$, and $\gamma_b$ is the color-spin degeneracy for parton $b$.

The collinear divergencies in the leading-order (LO) elastic scattering amplitudes $|{\cal M}_{ab\rightarrow cd}|^2$ are
regularized in the LBT model by a factor $S_2(\hat s, \hat t, \hat u) = \theta(\hat s\ge 2\mu_{D}^2)\theta(-\hat s+\mu_{D}^2\le \hat t\le -\mu_{D}^2)$,
 where $\hat s$, $\hat t$, and $\hat u$ are Mandelstam variables, and $ \mu_{D}^2 = (3/2)g^2 T^2$
 is the Debye screening mass with 3 quark flavors. The corresponding elastic cross sections are $d\sigma_{ab\rightarrow cd}/d\hat t=|{\cal M}_{ab\rightarrow cd}|^2/16\pi \hat s^2$. The fixed value of the strong coupling constant $\alpha_s=g^{2}/4\pi\approx 0.15$ is fitted to experimental data. The differential inclusive rates for gluon bremsstrahlung is assumed to follow that from the high-twist approach \cite{Wang:2001ifa},
\ba \la{induced}
\frac{d\Gamma_{a}^{\rm inel}}{dzd\ell_\perp^2}=\frac{6\alpha_sP_a(z)\ell_\perp^4}{\pi (\ell_\perp^2+z^2m^2)^4} \frac{p\cdot u}{p_0}\hat{q}_{a} (x)\sin^2\frac{\tau-\tau_i}{2\tau_f},
\ea
where $P_a(z)$ is the splitting function for parton $a$ to emit a gluon,  $a\rightarrow a+g$,  with momentum fraction $z$. $\ell_\perp$ is the transverse momentum of the emitted gluon, $m$ is the mass of the propagating parton, $\tau_f=2p_0z(1-z)/(\ell_\perp^2+z^2m^2)$ is the gluon formation time and $\tau_i$ is the time of the last gluon emission.
The elastic scattering rate in the inelastic processes has been factorized into the jet transport coefficient,
\begin{equation}
\hat{q}_{a}(x)=\sum_{bcd}\rho_{b}(x)\int d\hat t q_\perp^2 \frac{d\sigma_{ab\rightarrow cd}}{d\hat t},
\label{eq-qhat}
\end{equation}
which is defined as the transverse momentum transfer squared per mean-free-path in the local comoving frame of the QGP medium.  The local parton density $\rho_{b}(x)$ includes the degeneracy factor. A Poisson distribution with the mean value $\langle N^a_g \rangle=\Delta\tau\Gamma_a^{\rm inel}$ of gluon multiplicity is assumed to simulate multiple gluon radiations associated with each elastic scattering. Global energy and momentum conservation is ensured in each scattering with multiple gluon radiations.

For each scattering in the LBT model,  the initial thermal parton $b$ is recorded as a ``negative'' parton and they are also allowed to propagate in the medium according to the Boltzmann equation. The energy and momentum of these ``negative" partons are subtracted from all final observables to account for the back-reaction in the Boltzmann transport equations. They are part of the jet-induced medium response and manifest as the diffusion wake behind the wave of propagating jet shower partons.

\section{Jet energy loss and jet-induced medium response} 

The LBT model has been used to describe both single inclusive light and heavy flavor hadron suppression \cite{Cao:2017hhk}, $\gamma$-hadron \cite{Chen:2017zte}, single inclusive jet \cite{He:2018xjv}, $\gamma$-jet \cite{Luo:2018pto} and $Z^0$-jet correlations~\cite{Zhang:2018urd}.  In these studies of inclusive jets and $\gamma$-jets, initial jet production in $p+p$ collisions is simulated with PYTHIA. The jet shower partons are then transported through the QGP according to the LBT model with the bulk medium given by the CLVisc \cite{Pang:2018zzo} hydrodynamic simulations.

\begin{figure}[htbp]
\begin{center}
    \includegraphics[width=6.0cm]{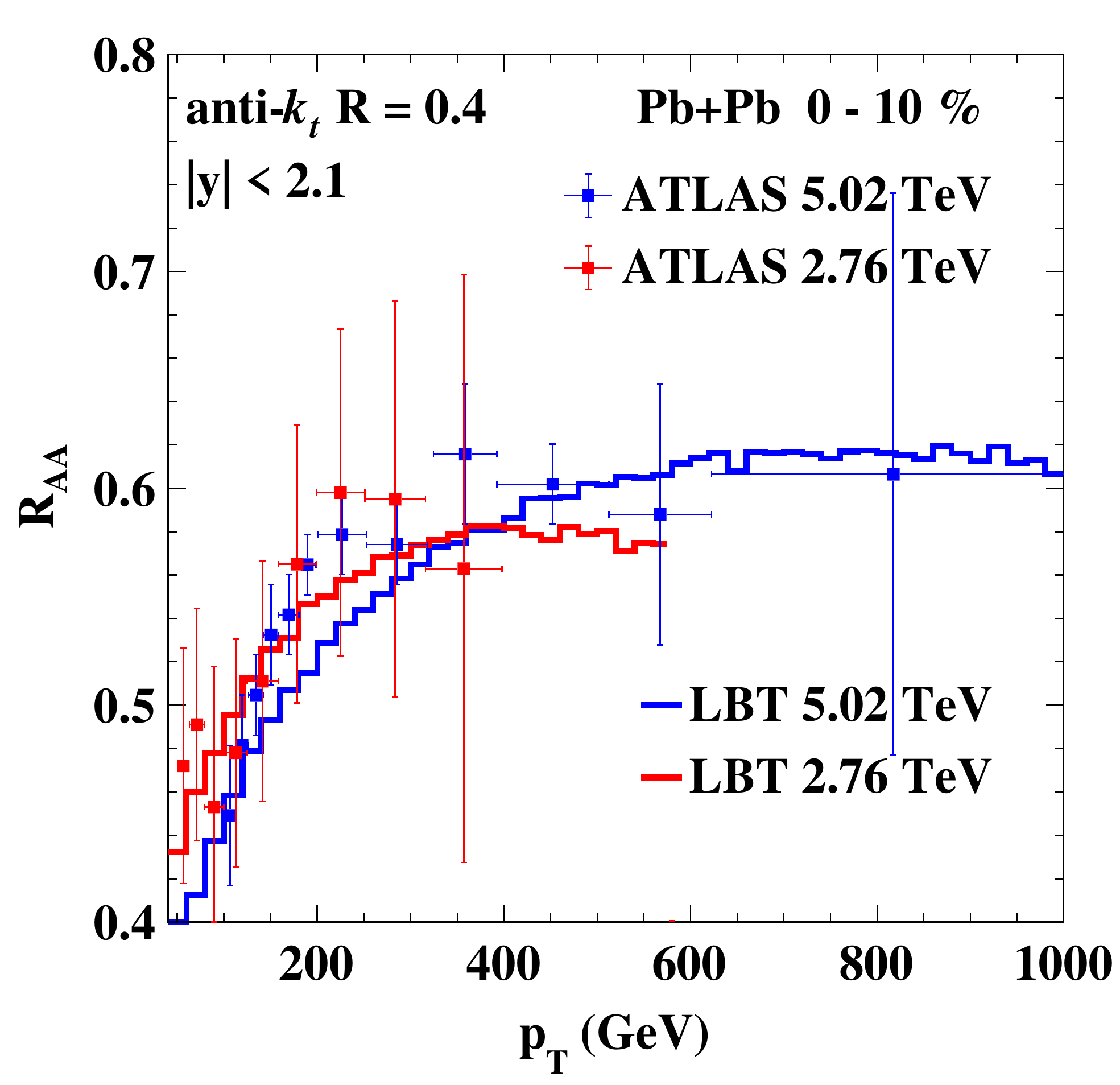}\hspace{0.2in}
    \includegraphics[width=6.0cm]{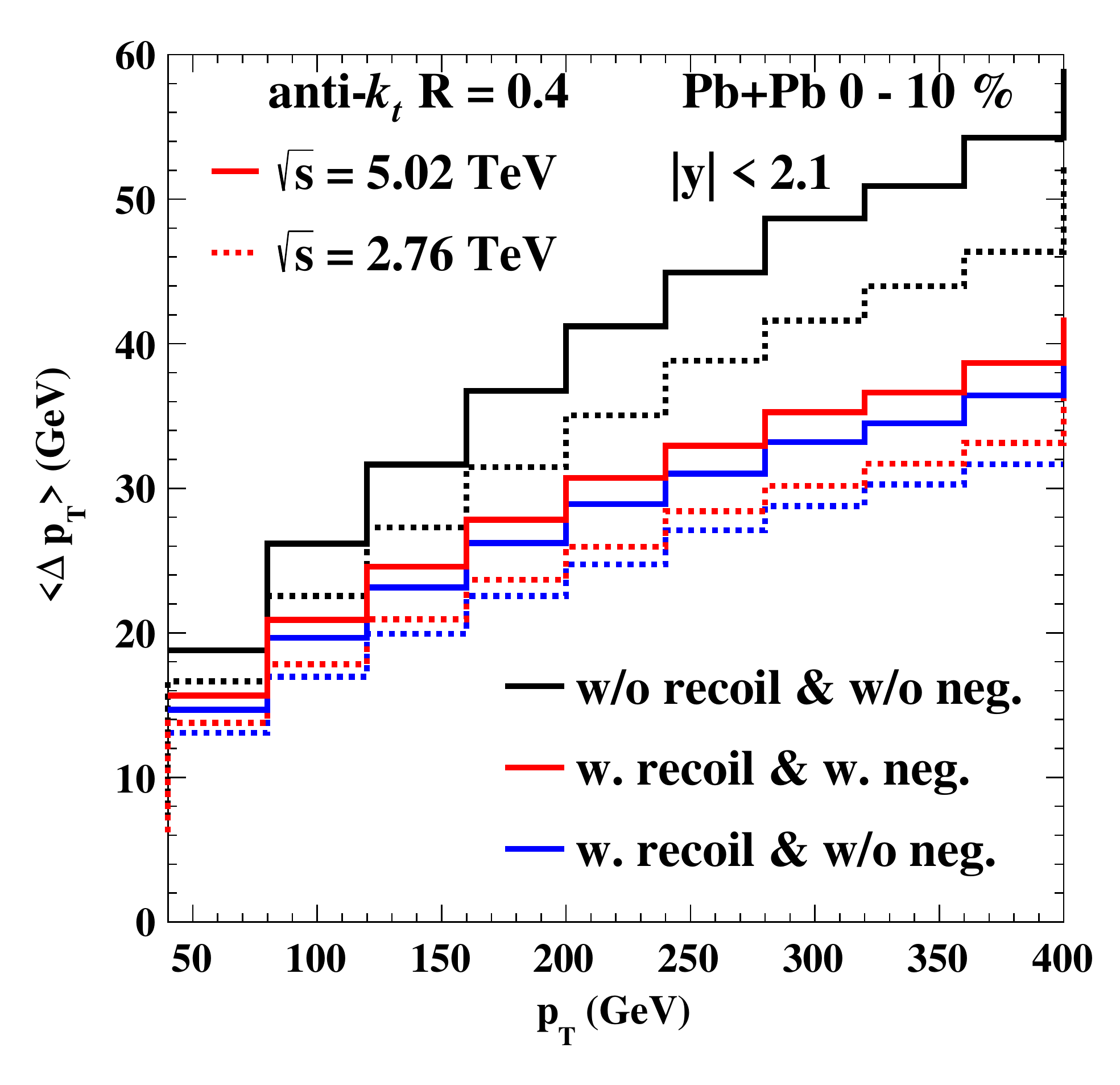}
    \end{center}
    \vspace{-0.3in}
    \caption{(left) LBT results on $R_{\rm AA}(p_T)$ for single inclusive jet spectra in 0-10\% central $Pb+Pb$ collisions at $\sqrt{s_{\rm NN}} = 2.76$ and 5.02 TeV as compared to ATLAS data \cite{Aad:2014bxa}. (right) Average jet transverse energy loss as a function of  jet $p_{T}$ in vacuum in the most central 10 \% $Pb+Pb$ collisions  at (solid) $\sqrt{s_{\rm NN}} = 5.02$ GeV  and  (dash) 2.76 TeV. See Ref.~ \cite{He:2018xjv} for details.}
    \label{RAA_twoEnergy}
\end{figure}

The suppression factors for single inclusive jet spectra in $Pb+Pb$ collisions at $\sqrt{s}=2.76$ and 5.02 TeV are shown in  Fig.~\ref{RAA_twoEnergy} (left) together with the latest  ATLAS data \cite{Aad:2014bxa}.  The LBT results are consistent with the experimental data and show very weak colliding energy dependence at LHC energies  despite the fact that the initial parton density at 5.02 TeV is about 20\% higher than at 2.76 TeV.  The single inclusive jet suppression factors also show very weak transverse momentum dependence in the range of the experimental coverage. These two features are the consequences of the initial parton spectra and the transverse momentum dependence of jet energy loss which is influenced by both jet-shower-medium interaction and the propagation of recoil partons \cite{He:2018xjv}.

To demonstrate the colliding energy and transverse momentum dependence of the jet energy loss, we show the averaged energy 
loss $\langle \Delta p_T\rangle$ in Fig.~\ref{RAA_twoEnergy} (right) for leading jets in the 0-10\% most central $Pb+Pb$ collisions at two colliding energies, $\sqrt{s_{\rm NN}}=2.76$ and 5.02 TeV, from LBT simulations. The difference between the final transverse energies of the vacuum and medium-modified leading jet is defined as the jet transverse energy loss as shown in Fig.~\ref{RAA_twoEnergy} (right)  as a function of the vacuum jet transverse energy.  The transverse jet energy loss at $\sqrt{s_{\rm NN}}=5.02$ TeV is indeed about 15\% larger than at $\sqrt{s_{\rm NN}}=2.76$ TeV in the $p_T=50-400$ GeV/$c$ range when the medium response (recoil and ``negative" partons) is taken into account.  It increases with the vacuum jet transverse energy logarithmically.  Such a weak $p_T$-dependence of the jet transverse energy loss is caused by a combination of effects due to jet-induced medium response, radial expansion and jet flavor (quarks and gluons) composition.

\begin{figure}[htbp]
\begin{center}
    \includegraphics[width=6.0cm]{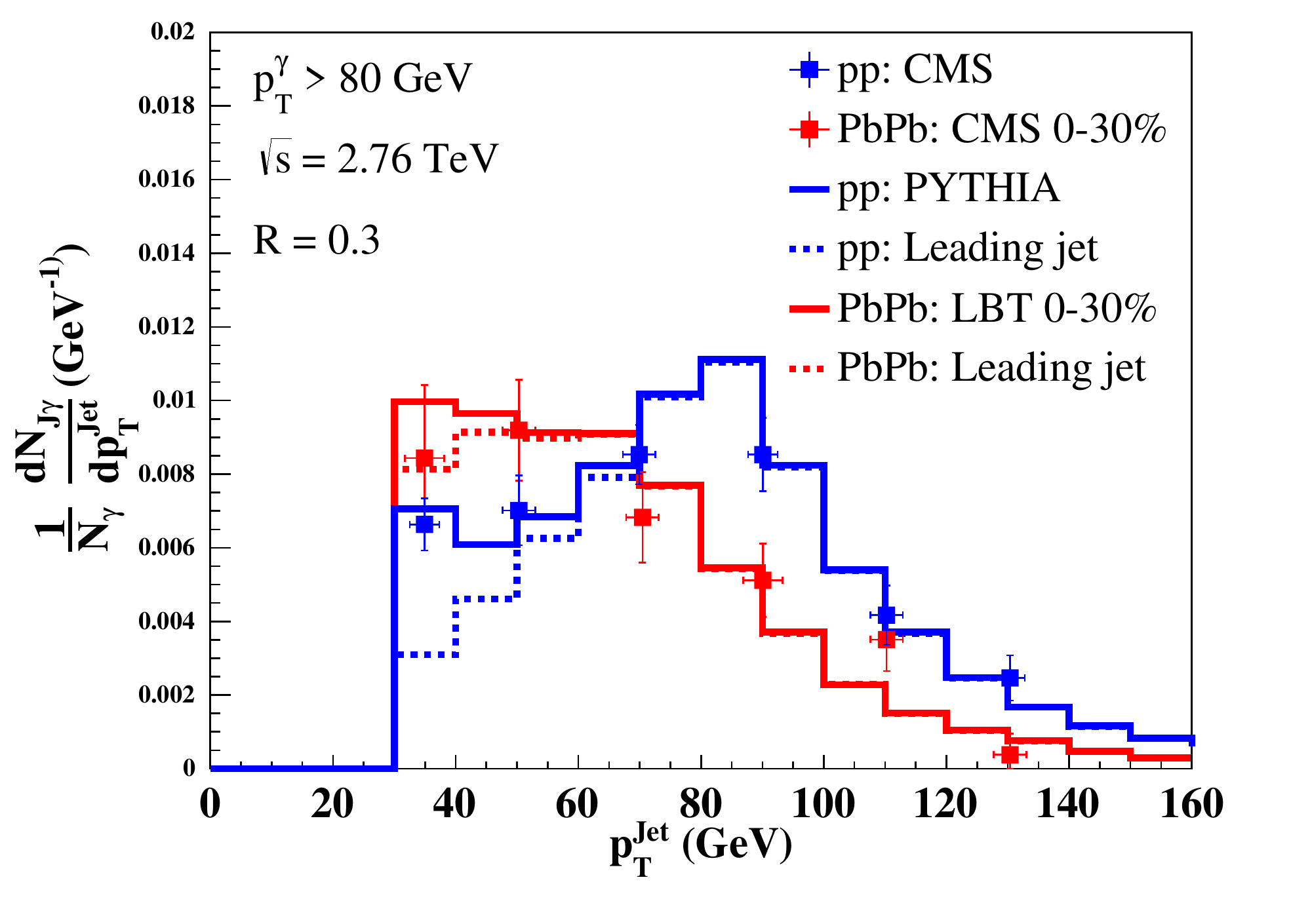}\hspace{0.2in}
    \includegraphics[width=6.0cm]{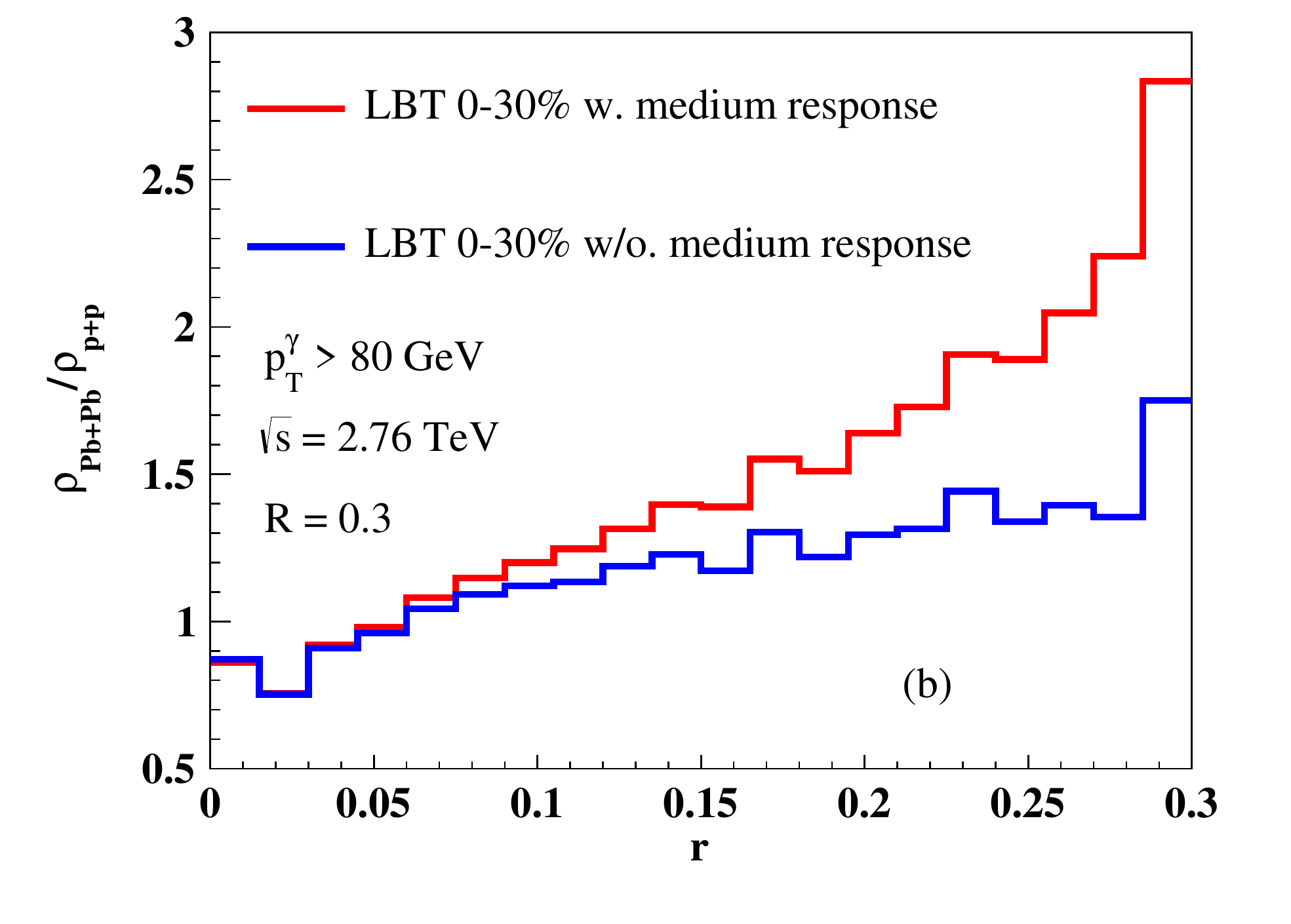}
    \end{center}
        \vspace{-0.3in}
    \caption{(left) LBT results on $\gamma$-jet spectra in $p+p$ and central $Pb+Pb$ collisions as compared to CMS data \cite{Chatrchyan:2012gt} and (right) the modification of jet transverse profile with and without medium response. See Ref.~cite{Luo:2018pto} for details.}
    \label{gammajet}
\end{figure}

The LBT can also explain CMS experimental data \cite{Chatrchyan:2012gt} on $\gamma$-jet distributions as shown in Fig.~\ref{gammajet} (left). For fixed $p_T^\gamma$, the peak of jet distributions, especially these of the leading jets (dashed lines) are shifted to smaller values of $p_T^{\rm jet}$ due to jet energy loss, which includes the effect of the medium recoil.  Some of these recoil medium partons should also fall within the jet cone and contribute to soft hadrons within a jet due the medium response. These soft hadrons will also modify the jet transverse profile.   Shown in Fig.~\ref{gammajet} (right) are the medium modification factors of the jet transverse profile with and without medium response. One can see that the jet transverse profile is significantly enhanced at large radius due to contributions from the medium response. These soft partons from medium recoil are also shown to lead to the enhancement of jet fragmentation functions at small momentum fraction $z$ \cite{Chen:2017zte}. The onset of the enhancement of soft hadrons is found to occur at a constant transverse momentum which reflects the thermal nature of hadrons from the medium response. 

\section{ Retrospect and Outlook}

\begin{figure}[htbp]
\begin{center}
    \includegraphics[width=6.0cm]{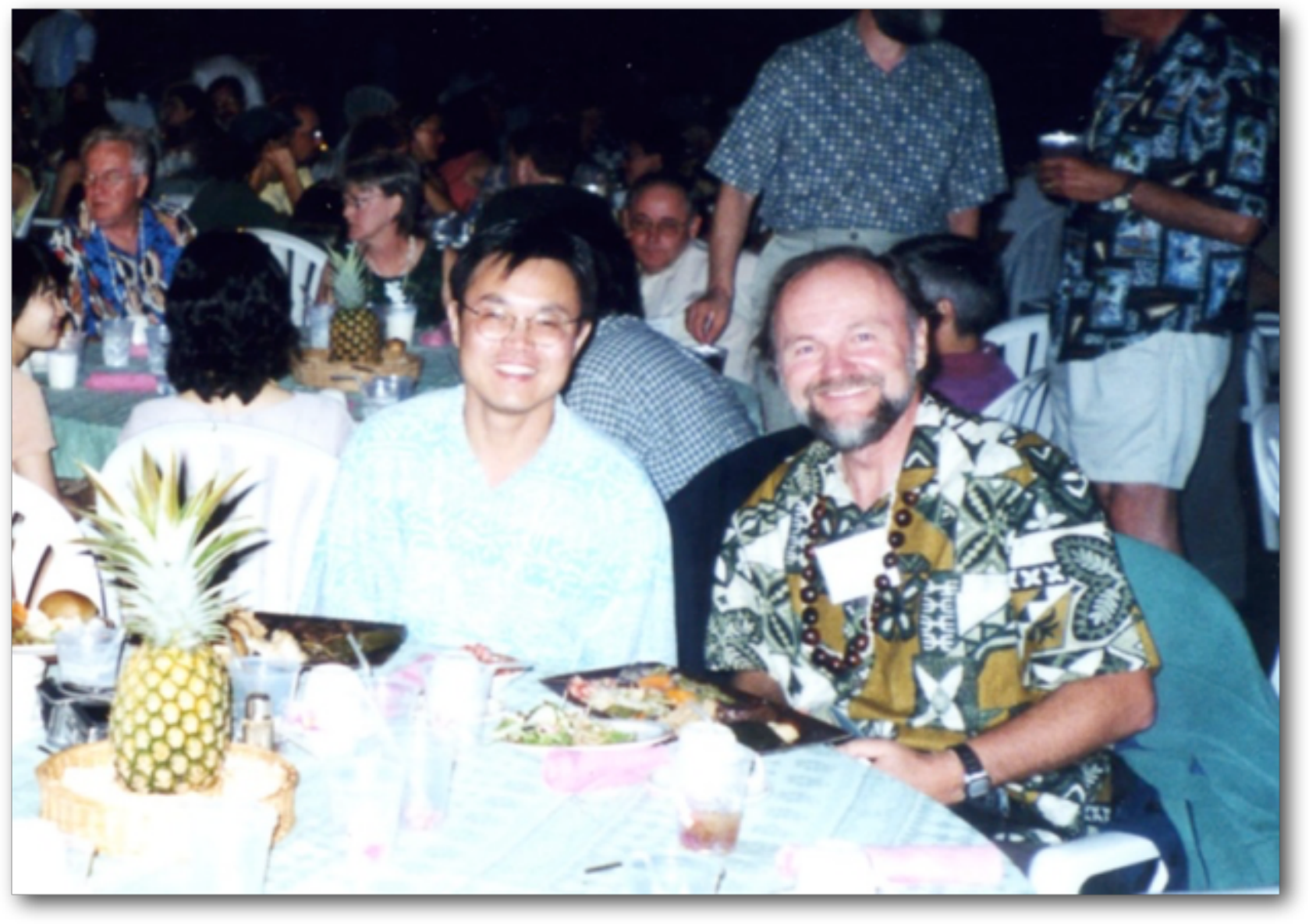}\hspace{0.5in}
    \includegraphics[width=5.9cm]{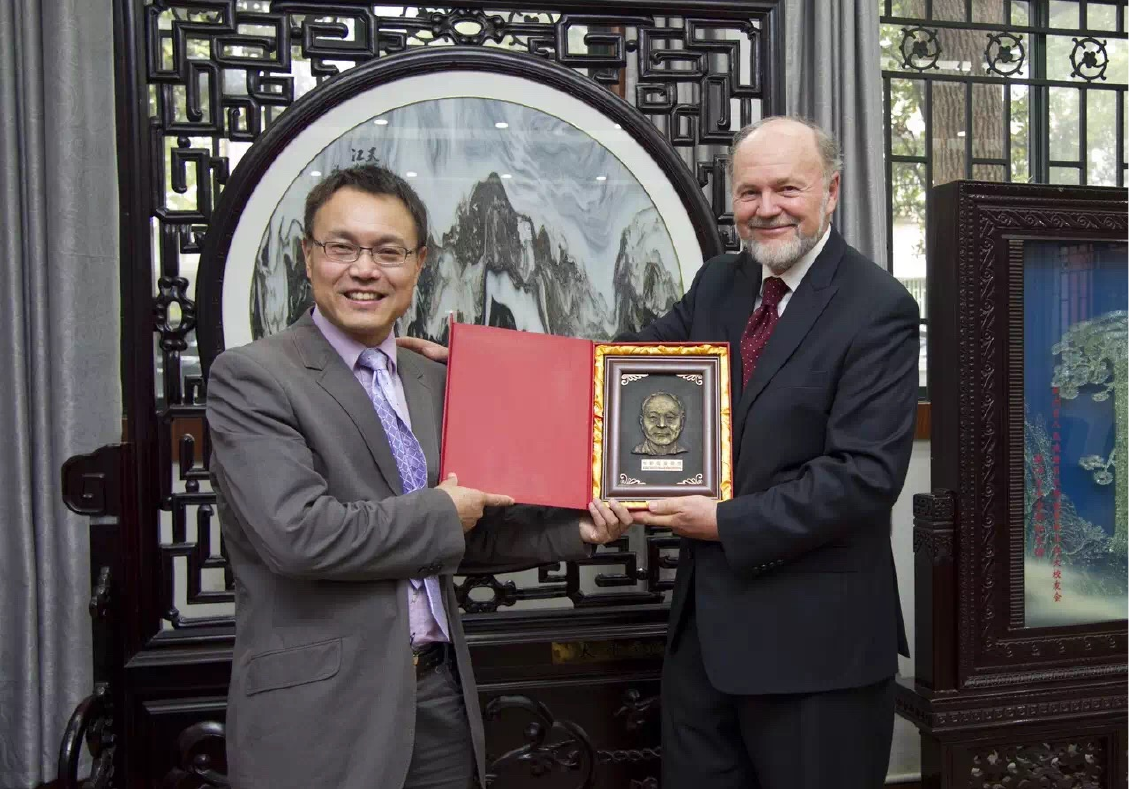}
    \end{center}
        \vspace{-0.3in}
    \caption{(left) Miklos and I together at the luau banquet during the joint APS-JPS DNP fall meeting in Maui, Hawaii on October 17-20, 2001. (right) Miklos and I at the award ceremony for his appointment as the Bian Peng Visiting Professor at CCNU on October 17, 2015.}
    \label{pic}
\end{figure}

The last 30 years have been an exciting time for a physicist working in the area of high-energy heavy-ion collisions, from the early years of RHIC construction and physics simulations, to the first data coming from RHIC detectors and  results from heavy-ion experiments at the LHC in the last decade. I  have been very fortunate to be in the right place and at the right time and got involved in this physics endeavor with thousands physicists worldwide. I feel blessed to have the opportunity to work and collaborate with Miklos and many other colleagues during this exciting period of a lifetime. It  has not only enriched my knowledge of our nature through the world of physics but also has given me the opportunity to enjoy my life as a physicist and a friend of so many people, Miklos in particular. Physics is fun as seen in the left picture in Fig.~\ref{pic} we took together at the joint APS-JPS DNP fall meeting in Maui, Hawaii on October 17-20, 2001.  Miklos has returned to physics full time at LBNL since his retirement from Columbia University and was also appointed as the Bian Peng Visiting Professor at CCNU (see the right picture in Fig.~\ref{pic} taken during the award ceremony at CCNU on October 17, 2015). We will have many more years to collaborate.

After 30 years of theoretical development and experimental studies at both RHIC and LHC, jet quenching has become a powerful tool as a tomographic probe of the QGP in high-energy heavy-ion collisions. We have entered a stage for more precision studies of the physics properties of QGP using jets. More theoretical investigations are needed in order to improve the precision via inclusion of  higher order correlations, color coherence and de-coherence and multiple gluon emissions with overlapping formation time. Understanding of jet-induced medium response is still in its nascent stage. Its full potential to probe the transport response of QGP  is still to be explored. 

\section*{Acknowledgements}

This work is supported by DOE under Contract No. DE-AC02-05CH11231,  by NSF under grant No. ACI-1550228 within the JETSCAPE Collaboration, by NSFC under Grants No. 11890714 and No. 11861131009.

\end{document}